# Symmetry transformations in Batalin-Vilkovisky formalism.

Albert Schwarz*
Department of Mathematics, University of California,
Davis, CA 95616
ASSCHWARZ@UCDAVIS.EDU

**Abstract**

Let us suppose that the functional $S$ on an odd symplectic manifold satisfies the quantum master equation $\Delta_\rho e^S = 0$. We prove that in some sense every quantum observable (i.e. every function $H$ obeying $\Delta_\rho(He^S) = 0$) determines a symmetry of the theory with the action functional $S$.

This short note was inspired by the paper [2] where the results of [1] were applied to obtain the description of the gauge transformations in Batalin-Vilkovisky theory. We begin with the observation that [1] contains conditions of physical equivalence of different solutions to the master equivalence and use these conditions to give a very transparent analysis of symmetry transformations in BV-approach. Let us recall some notions and results of [1].

Let us fix a manifold $M$ provided with odd symplectic structure ($P$-structure). Let us suppose that the volume element in $M$ is specified by the density $\rho$. We say that this volume element determines an $SP$-structure in $M$ if $\Delta_\rho^2 = 0$. (Here the operator $\Delta_\rho$ acts by the formula

$$\Delta_\rho A = \frac{1}{2} div_\rho K_A, \qquad (1)$$

where $K_A$ denotes the hamiltonian vector field corresponding to $A$ and the divergence $div_\rho$ is calculated with respect to the density $\rho$.) By definition a function $S$ satisfies the quantum master equation on an $SP$-manifold $M$ if

$$\Delta_\rho e^S = 0. \qquad (2)$$

Such a function $S$ can be considered as an action functional and determines physical quantitional by means of integration over Lagrandian submanifolds of $M$.

*Research supported in part by NSF grant No. DMS-9201366

The following statement was proven in [1](see Lemma 4 and Eqn.(32)):
The density $\tilde{\rho} = e^\sigma \rho$ determines a new $SP$-structure in $M$ if and only if

$$\Delta_\rho \sigma + \frac{1}{4}\{\sigma, \sigma\} = 0. \tag{3}$$

If $S$ is a solution to the quantum master equation(1) then $\tilde{S} = S - \frac{1}{2}\sigma$ satisfies the quantum master equation

$$\Delta_{\tilde{\rho}} e^{\tilde{S}} = 0 \tag{4}$$

corresponding to the new $SP$-structure. The action functional $\tilde{S} = S - \frac{1}{2}\sigma$ on the manifold $M$ with the new $SP$-structure describes the same physics as the action functional $S$ on the manifold $M$ with the old $SP$-structure. In particular

$$\int_L e^{\tilde{S}} d\tilde{\lambda} = \int_L e^S d\lambda \tag{5}$$

for every Lagrangian submanifold $L \subset M$. (Here $d\lambda$ and $d\tilde{\lambda}$ denote the volume elements on $L$ determined by new and old $SP$-structure correspondingly.)

One can introduce the notion of quantum observable for the theory with the action $S$ on an $SP$-manifold $M$ in the following way. We will say that an even function $A$ determines a quantum observable if

$$\Delta_\rho(A e^S) = 0 \tag{6}$$

or, equivalently, if $\Delta_\rho A + \{A, S\} = 0$. It is easy to check that for every observable $A$ we have also $\Delta_{\tilde{\rho}}(A e^{\tilde{S}}) = 0$; in other words the quantum observables for the action functionals $S$ and $\tilde{S}$ coincide. Moreover

$$\int_L A e^{\tilde{S}} d\tilde{\lambda} = \int_L A e^S d\lambda \tag{7}$$

for every observable $A$ and for every Lagrangian manifold $L$. This equation can be considered as a little bit more precise expression of physical equivalence of action functionals $S$ and $\tilde{S}$ than (5). The condition (6) is equivalent to the requirement that $S + \varepsilon A$ is a solution to the quantum master equation for infinitesimal $\varepsilon$; therefore we can derive (7) applying (4) to the functionals $S + \varepsilon A$, $\tilde{S} + \varepsilon A$ where $\varepsilon \to 0$.

Let us use the statement above in the case $\sigma = 2S$. As was mentioned in [1] the equation (2) can be represented in the form $\Delta_\rho e^{\sigma/2} = 0$; therefore it is satisfied for $\sigma = 2S$. We arrive to the following conclusion:

The theory with the action $\rho$ is physically equivalent to the theory with the trivial action functional $\tilde{S} = 0$ and the $SP$-structure determined by the density $\tilde{\rho} = e^{2S}\rho$.

The statement above can be used to analyze the symmetry transformations in Batalin-Vilkovisky approach. If $\tilde{S} = 0$ then the symmetries can be characterized as transformations of $M$ preserving the symplectic structure and the density $\tilde{\rho} = e^{2S}\rho$. Infinitesimal symmetry transformation correspond therefore to hamiltonian vector fields with zero divergence with respect to $\tilde{\rho}$, i.e. to functions $H$ obeying $\Delta_{\tilde{\rho}} H = 0$. One can say therefore that infinitesimal symmetry transformations are in one-to-one correspondence with quantum observables of our theory. It is important to emphasize that the notion of observable does not change when we replace $S$ by $\tilde{S} = 0$. Therefore we can describe symmetries also in terms of the original action $S$ and the original measure $d\mu = \rho dx$. (This description follows immediately from the description of symmetry transformations in the formulation with $\tilde{S} = 0$; one can prove it directly using (5),(6),(7).)

*We obtain that every function $H$ satisfying $\Delta_\rho H + \{H, S\} = 0$ (every quantum observable) determines a symmetry in the following sense. Neither the action functional $S$, nor the density $\rho$ are invariant with respect to an infinitesimal transformation with the Hamiltonian $H$, however the new action functional*

$$\tilde{S} = S + \varepsilon\{H, S\} \tag{8}$$

*and the new measure*

$$d\tilde{\mu} = d\mu(1 + 2\varepsilon\Delta_\rho H \tag{9}$$

*describe the same physics as the old action functional $S$ and the old measure $d\mu = \rho dx$.*

Here $\varepsilon$ is an infinitesimal parameter. The formula (9) follows immediately from (1) and from the standard formula

$$d\tilde{\mu} = d\mu + \varepsilon d\mu \ div_\rho K \tag{10}$$

for the variation of measure by the infinitesimal transformation $x \to x + \varepsilon K$, where $K$ is an arbitrary vector field.

It is essential to stress that quantum observables don't remain intact by the symmetry transformation (8),(9). Namely, one should replace a quantum observable $A$ by the observable $\tilde{A} = A + \varepsilon\{H, A\}$.